\documentclass[letterpaper, 10 pt, conference]{ieeeconf}  

\IEEEoverridecommandlockouts                              

\usepackage{graphicx} 
\usepackage{bm}

\usepackage{bbding}
\usepackage{physics}
\usepackage{amsmath}
\usepackage{subcaption} 
\usepackage{caption}
\usepackage[
backend=biber,
style=numeric-comp,
sorting=none
]{biblatex}
\title{\LARGE \bf
3D Dynamics of a Premagnetized Gas-puff Z-pinch implosion
}

\author{P. Phillips$^{1, ^*}$, M. Escalona$^{1}$, P. Retamales$^{1}$, M. Ribeiro$^{1}$, F. Veloso$^{1}$, J. C. Valenzuela$^{1,\dagger}$
\thanks{}
\thanks{* pelayo.phillips@uc.cl, $\dagger$ jcvalen1@uc.cl}%
\thanks{$^{1}$Instituto de Física, Pontificia Universidad Católica de Chile, Santiago 7820436, Chile}%
}

\begin{document}

\thispagestyle{plain}
\pagestyle{plain}
\maketitle

\begin{abstract}

Gas-puff Z-pinches are pulsed-power-driven plasma implosions used as compact laboratory sources of soft X-rays and fusion neutrons and as a platform for magneto-inertial fusion research. When a preembedded axial magnetic field is applied, the implosion becomes intrinsically three-dimensional, with radial compression, self-generated rotation, and (in the presence of zippering) axial flow all contributing to the dynamics and to the energy balance of the pinch. The simultaneous resolution of all three velocity components is therefore essential for a complete description. We present the first simultaneous, spatially-resolved measurement of all three velocity components (radial, azimuthal, and axial) in an annular magnetized argon gas-puff Z-pinch, performed using Collective Thomson Scattering along three orthogonal lines of sight from the same scattering volume at each time step. Measurements were carried out on the Llamp{\"u}dke{\~n} pulse-power generator (400 kA peak, 200 ns rise time), for applied axial fields ranging from 0.04 to 0.26 T using two coil configurations: a double coil with negligible initial radial field at the probed plane ($z = 8$ mm), and a single coil that imposes a finite initial radial field. Three principal results are reported. First, the axial velocity component, which had not previously been measured experimentally in this configuration, reaches 60--70 km\,s$^{-1}$ near the axis at low applied fields ($B_{z0} < 0.1$ T) and is suppressed to within $\pm 20$ km\,s$^{-1}$ for stronger applied fields, in correlation with the reduction of the zippering angle, with direct implications for the implosion energy balance. Second, the self-generated rotation extends across the full plasma diameter, not only at the periphery, and the diametrical profile of the azimuthal velocity decreases toward the axis with an exponential-like shape consistent with the underlying current density distribution; this feature was not visible in previous edge-localized measurements. Third, rotation persists in the double-coil case ($B_{r0} \approx 0$) and is enhanced in the single-coil case (finite $B_{r0}$), supporting the interpretation that $B_r$ develops self-consistently during the implosion and drives the rotation through a $J_z \times B_r$ torque. These results constrain the role of each magnetic-field component and motivate direct measurement of $B_r$ and the current density distribution as the next step.

\end{abstract}


\section{\label{sec:level1}Introduction}

The magnetized gas-puff configuration has been studied extensively in the past due to its positive effect in mitigating instabilities commonly found in Z-pinch plasmas, such as Magneto-Rayleigh-Taylor (MRT) instabilities \cite{narkis2021, conti2020}, and because of its importance in magneto-inertial fusion (MIF) research \cite{slutz2012, klir2015, coverdale2007}. 

In this configuration, an axial magnetic field is applied across the anode–cathode (A–K) gap of the generator prior to plasma formation. This field is typically generated by one or two coils driven by a current pulse with a rise time much longer than the plasma lifetime \cite{Felber1987, chaikovsky2003, mikitchuk2014}. After the magnetic field has diffused into the electrodes, a gas injector allows the flow of a gas in the shape of a uniform cylinder through the A-K gap \cite{valenzuela2017, shiloh1978}. Then, a fast, high-amplitude pulse (typically between 300 kA-2 MA and 100-300 ns in duration) is applied, producing a plasma that rapidly implodes toward its axis due to the magnetic pressure associated with the self-generated azimuthal magnetic field \cite{giuliani2015}. Given that the implosion timescale is much shorter than the current pulse used to establish the axial magnetic field, this current remains constant during the implosion, and the axial field is governed mostly by the interplay between resistive diffusion and advection with the flow.

An important motivator behind the study of this configuration is its relevance to magneto-inertial fusion research and, more specifically, to the physics of helical instabilities and magnetic flux compression that are common to many axially magnetized Z-pinches, including MagLIF \cite{awe2014}. The high repetition rate of a gas-puff makes it a practical platform for parametric studies of these phenomena. One example of this has to do with the study of helical instabilities that are found, not only on magnetized gas-puffs \cite{chen2024}, but also in many other Z-pinch configurations that use axial fields \cite{yager2016, atoyan2016, yager2018}, that have been fundamental in the study of some of the mechanisms behind the formation and development of these structures in MagLIF implosions \cite{awe2014}.

Recently it was discovered that the application of a preembedded axial magnetic field in a gas-puff induces self-generated rotations in the plasma \cite{cvejic2022}, where azimuthal velocities at the outer edge of the plasma were measured at different axial points using Doppler spectroscopy. It was shown that these rotations could reach velocities of the same magnitude as the peak implosion velocity of the plasma. These results have very important implications for our understanding of Z-pinches and inertial fusion research, since this impacts the stagnation dynamics, and it also could explain several experimental phenomena observed in Z-pinches \cite{cvejic2022}.

Additionally, the energy balance of the gas-puff was recently discussed in detail \cite{Lavine2024}, where a triple nozzle gas-puff was studied with and without the presence of an axial magnetic field using Thomson Scattering (TS). Interestingly, in contrast with what was reported previously \cite{cvejic2022}, where the radial and azimuthal velocities were fairly similar during the implosion, here the azimuthal velocity reported when using an axial field was fairly small compared to the radial velocity until the stagnation phase of the pinch, where the azimuthal velocity becomes considerably larger \cite{Lavine2024}. It was also reported that angular momentum was conserved in the middle of the A-K gap, which contrasts somewhat with what was outlined previously in another work \cite{cvejic2022}, where the azimuthal velocity in the middle of the A-K gap was constant during the implosion, and where near the electrodes the change in the azimuthal velocity of the plasma was consistent with isorotation. In that work, $v_\theta$ increased near the anode and decreased near the cathode as the implosion progressed. Because of these different results, some questions are still unanswered as to how reproducible or universal the conditions that produce these rotations are, and if the isorotation previously reported \cite{cvejic2022} is maintained under different conditions and cathode geometries.

These different results also raise questions about the impact on the energy balance that this mechanism has under these different conditions. Another aspect of the gas-puff dynamics that is often ignored, but can impact the energy balance, is the axial velocity. This velocity can develop because of zippering and can account for a substantial fraction of the plasma kinetic energy, as suggested by simulations \cite{deeney1993}. This aspect of the implosion could explain some of the discrepancies found between the coupled energy and the energy of the pinch near the stagnation measured on previous experimental measurements \cite{Lavine2024}, and therefore it is important to study it and the impact that the axial field can have on it. 

In this paper, we present the first simultaneous spatially-resolved measurement of the three velocity components in a magnetized argon gas-puff Z-pinch implosion. A Thomson Scattering (TS) diagnostic with three orthogonal lines of sight was implemented, allowing $v_r$, $v_\theta$, and $v_z$ to be reconstructed from the same scattering volume at each time step. Measurements were performed for applied axial fields ranging from 0.04 to 0.26 T using two coil geometries: a double-coil configuration, in which the initial $B_{r0}$ at the probed plane ($z = 8$ mm) is close to 0 mT, and a single-coil configuration, in which a finite $B_{r0}$ is imposed by design (we use 'double-coil' rather than 'Helmholtz pair' because the coil spacing is approximately half a coil radius, not one radius). Three principal results emerge from this dataset. First, we report the first experimental characterization of the axial velocity in a magnetized gas-puff: $v_z$ reaches 60--70~km\,s$^{-1}$ near the axis at low applied fields ($B_{z0} < 0.1$ T) and is suppressed to within $\pm 20$~km\,s$^{-1}$ at higher fields, in correlation with the reduction of the zippering angle. Second, the self-generated rotation is not limited to the periphery of the plasma: the diametrical profile of $v_\theta$ measured here decreases toward the axis with an exponential-like shape consistent with the underlying current density --- a feature that was not visible in previous edge-localized measurements~\cite{cvejic2022, Lavine2024}. Third, by comparing the double-coil and single-coil cases, we isolate the role of the seed radial field: rotation persists when $B_{r0} \approx 0$ and is enhanced when a finite $B_{r0}$ is imposed, supporting the interpretation that $B_r$ develops self-consistently during the implosion and that the rotation is driven by a $J_z\times B_r$ torque. We additionally discuss the effect of the applied field on the radial velocity, which exhibits a qualitative change in the stagnation timing above $B_{z0} \approx 0.26$~T that warrants further investigation.

\section{Experimental Setup}

\begin{figure}[tb]
    \centering
    \includegraphics[width=0.48\textwidth]{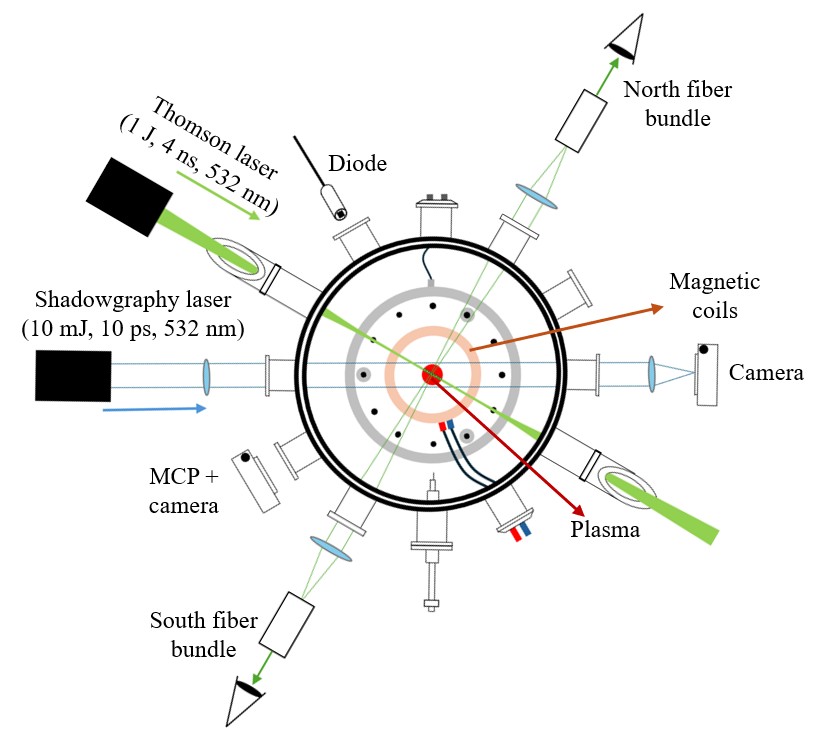}
    \caption{Diagram of the vacuum chamber and the diagnostics used in this work.}
    \label{fig:diagrmaCamara}
\end{figure}

\begin{figure*}[t!]
    \centering
    \includegraphics[width=0.95\textwidth]{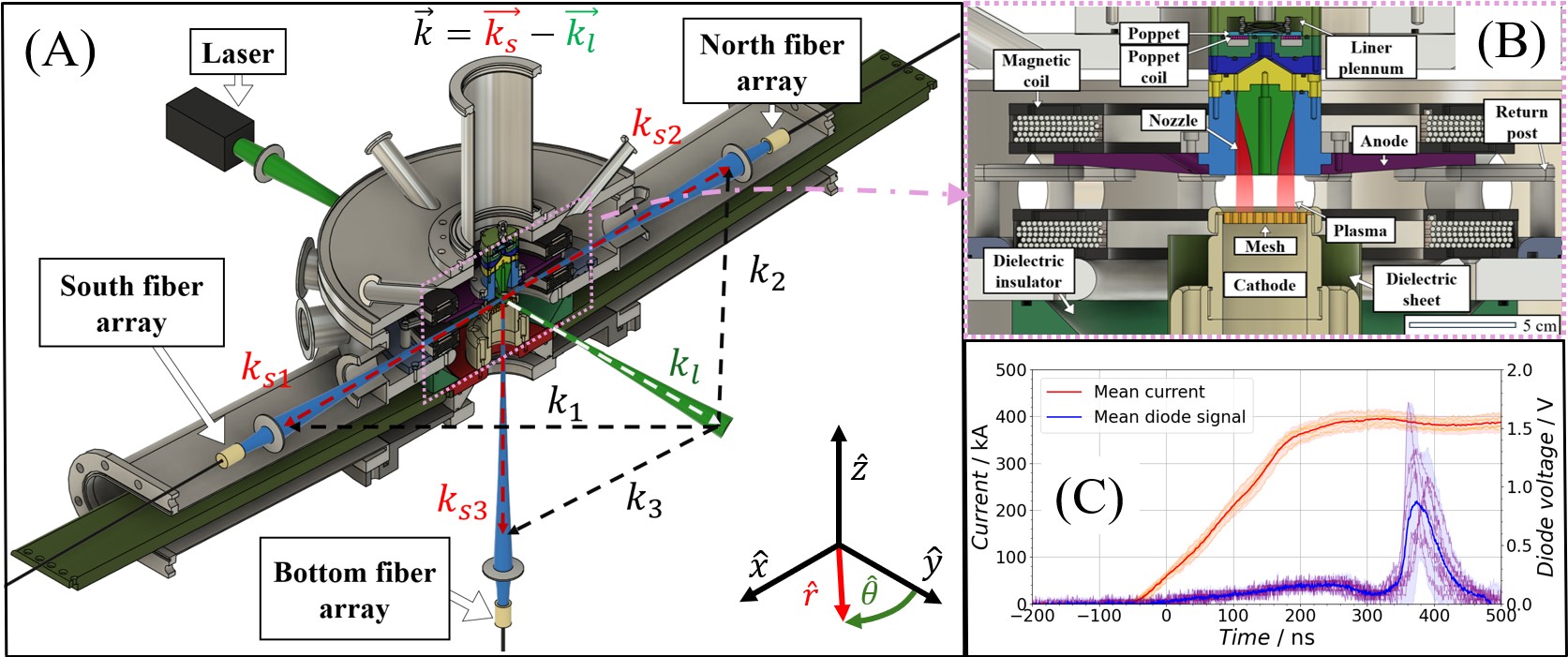}
    \caption{A) View of the Thomson Scattering configuration inside the chamber. B) Cross section of the injector inside the chamber. C) An example of the currents generated by Llampüdkeñ and the x-ray signals measured by the diode for the shots where the applied field was $B_{z0}=0.04$ T.  Here we see that the current peaks around 330-340 ns, and that after that the current in all cases is sustained due to current crowbarring. In this case the 0 in the time axis is defined as the moment where the current reaches 10\% of its peak value.}
    \label{fig:Chamber}
\end{figure*}

The measurements presented in this article were taken in Llampüdkeñ \cite{chuaqui1999}, a Marx-bank-based pulse-power generator capable of delivering a peak current of 400 kA in 200 ns when charging each of the capacitors to 22 kV. The pulse is delivered to an annular cylindrical gas load of argon. The gas is injected via a Gas-puff system, prefilled to 2 bar, with an initial outer radius of 14 mm and an inner radius of 6 mm, resulting in a linear density of 15 $\mu$g/cm \cite{Julio2023}. The relative position of each diagnostic can be seen in Fig. \ref{fig:diagrmaCamara}, where we see a diagram of the chamber looking from the top, along with the diagnostics used. When operating the gas injector, a strong current is applied to a copper coil that lifts an aluminum alloy plate that separates the plenum and the nozzle, and allows the gas to flow \cite{valenzuela2017}. A breakdown pin is placed in the nozzle that detects the gas flowing and sends a signal that triggers the generator. The gas is collimated by the nozzle and exits the injector at supersonic speeds as a uniform hollow cylinder that flows through the 16 mm A-K gap. Fig. \ref{fig:Chamber}B shows the inside of the vacuum chamber with the components described, and Fig. \ref{fig:Chamber}C shows some current traces produced by Llampüdkeñ. Fig. \ref{fig:Chamber}C also shows various examples of the x-ray signal measured by an AXUV5 diode, which is fitted with a 10~$\mu$m beryllium filter and is sensitive to radiation from 0.0124 to 1 nm.

For the double-coil configuration, the axial magnetic field was generated by 2 parallel coils, each with an outer radius of 120 mm and inner radius of 75 mm, placed on the A-K gap separated by 26 mm, as can be seen in Fig. \ref{fig:Chamber}B. In order to ensure that the magnetic field was axially and radially uniform, the coils were first designed using CAD software and then ANSYS Maxwell simulations were used to ensure the uniformity of the field produced. After the coils were fabricated, B-dot probes were used to characterize the field, which showed a good agreement with the ANSYS simulations. For the single-coil configuration the same process was applied, and in this case the bottom coil shown in Fig. \ref{fig:Chamber}B placed around the cathode was the only one used to generate the field.

Thomson Scattering was implemented through a Q-switched 1J 532nm Nd:YAG probe laser (EKSPLA NL310) with a pulse width of 4 ns. The laser's polarization was rotated to 45° (with respect to the horizontal) using a $\lambda/2$ waveplate, and then the laser passed through a 1500 mm converging lens and it entered the chamber through a Brewster window which was also rotated by 45° (with respect to the horizontal) in order to maximize the laser energy that enters the chamber. After this, the laser beam passed through the midpoint of the column at a height of 8 mm from the cathode (A-K gap of 16 mm), and the scattered light was collected and transmitted to the spectrometers. The axial velocity of a magnetized Gas-puff has yet to be experimentally measured, and some simulations suggest that it can be fairly large \cite{seyler2020, kumar2009}. Therefore, in order to accurately measure the dynamics and the energetics of the pinch, 3 bundles of fibers positioned at 90° from the probing laser were used; 2 bundles positioned in the geographical north and south of the chamber looking at the plasma diametrically on opposite sides, and another bundle placed at the bottom of the chamber looking under the plasma in the axial direction. Each of these bundles consists of 25 linearly arranged fibers of 200 $\mu$m in diameter, each separated by 300 $\mu$m, and all bundles collected the scattered laser light using a convex achromatic lens (100 mm focal length, 2.1 magnification) and a linear polarizer. The laser's polarization described earlier had to be rotated 45° in order to ensure that all bundles receive the same signal strength \cite{froula}.

The light collected by the north and south fiber bundles is then transmitted to a 500 mm focal length spectrometer (SpectraPro HRS-500), that has a 50 $\mu$m entrance slit and a 2400 lines/mm grating, coupled to a time-gated ICCD (Stanford 4Picos) which uses a time integration of 3 ns. The light collected by the axial/bottom fiber bundle is coupled to a 750 mm focal length spectrometer (Andor Shamrock 750), that has a 50 $\mu$m entrance slit and a 1800 lines/mm grating, and which is coupled to an ICCD (Andor iStar) that uses a time integration of 3 ns. The SpectraPro spectrometer has a spectral resolution of 0.06 nm, and the Andor spectrometer has a spectral resolution of 0.04 nm. Fig. \ref{fig:Chamber}A shows a representation of the relative positioning of the fiber bundles with respect to the TS laser, and it shows the direction of the wave vector. 

The velocity measured by TS ($v_k$) is the projection of the flow velocity of the probed volume of plasma into the TS vector $\vec{k}=\vec{k_s}-\vec{k_l}$, where $\vec{k_s}$ is the wavenumber of the scattered laser, and $\vec{k_l}$ is the wavenumber of the incident laser (see Fig. \ref{fig:Chamber}A). Therefore, the velocity measured by the north and south bundles will be given by the projection of the radial and azimuthal velocities into $\vec{k}$, or $v_{n} = v_r  \cos{\theta} + v_\theta\sin{\theta}$ and $v_{s} = v_r\cos{\theta}-v_\theta\sin{\theta}$ (here $\theta$ is defined with respect to the y-axis, see Fig. \ref{fig:Chamber}A), with $\theta \approx \pi/4$, and the light collected by the bundle located in the bottom will be given by $v_{b} = v_z \cos{\theta} - v_{r}\cos{\theta}$, the velocity components are: 

\begin{align}
    v_{r} &= (v_{n} + v_{s})/\sqrt{2} \label{vr} \\
    v_{\theta} &= (v_{n} - v_{s})/\sqrt{2} \label{vt} \\
    v_{z} &= -\sqrt{2} \cdot v_{b} + (v_{n} + v_{s})/\sqrt{2} 
    \label{vz}
\end{align}

In this frame of reference, $v_r>0$ is defined as a movement outwards from the center of the plasma, $v_\theta>0$ is defined as the clockwise movement, and $v_z>0$ is an upwards movement from the cathode ($z=0$ mm) to the anode ($z=16$ mm).

In order to be able to collect light with the bottom fiber, it was necessary to cut a symmetrical cross pattern in the cathode mesh so that the scattered light could be collected by the bottom fiber bundle. This cross has a width of 2~mm and a length of 15 mm, and it is aligned with the laser path. 

\begin{figure}
    \centering
    \includegraphics[width=0.48\textwidth]{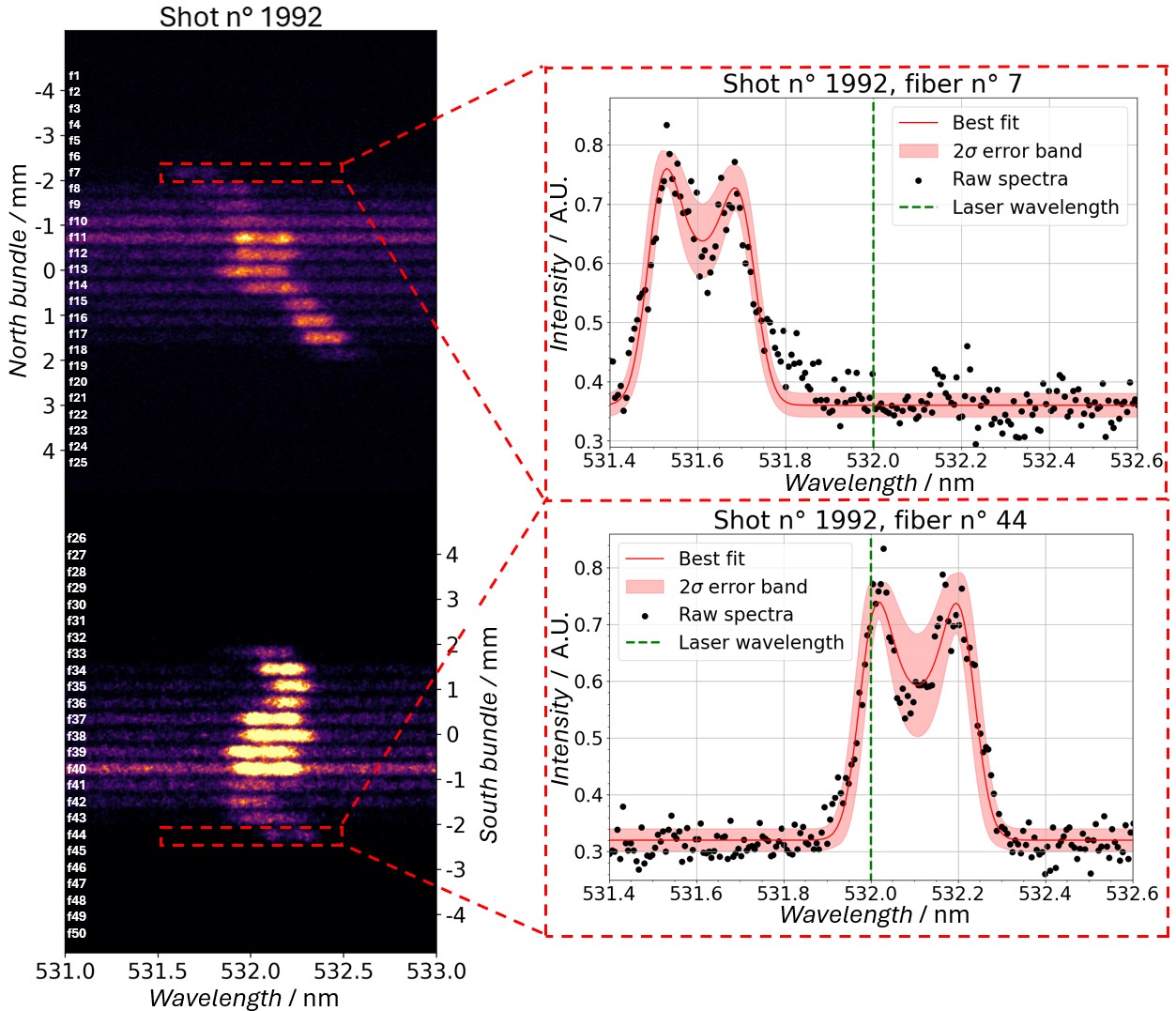}
    \caption{Image of the spectra collected by the north and south fiber bundles taken by the ICCD -18 ns before stagnation for shot n° 1992 using an initial field of 0.19 T. In the right side is the spectrum measured by the fibers looking at the same volume, where there is a clear difference in $v_k$ between the two fibers caused by the rotation of the plasma that red-shifts the spectrum taken by one fiber and blue-shifts the spectrum taken by the opposite fiber.}
    \label{fig:thomson1992}
\end{figure}

Before analyzing the TS spectra obtained, first it is necessary to identify if the scattering is collective or not, and also it is important to identify whether or not the electrons and ions in the plasma are magnetized, as this can heavily impact the spectral density function \cite{froula}. In order to identify if the TS spectra obtained are in the collective regime or not, we need to look at the $\alpha = \frac{1}{k\lambda_D}$ parameter, which relates the Debye length of the plasma with the scattering wave vector. The TS regime agrees mostly with the collective regime, with $\alpha \approx 0.5 - 2.6$ across the dataset and most spectra well above $\alpha = 1$. For the ions and electrons to be magnetized, or rather, for the spectral density function to be influenced by the magnetic field, first the scattering length needs to be much larger than the cyclotron radius of the particles. Second, the particles need to complete an appreciable fraction of a cyclotron orbit during the integration time of the spectrometer and the coherence time of the laser source \cite{froula}. The first and second conditions are respectively:

\begin{align}
    \rho_{e,i} &\ll L\\
    2\pi/\Omega_{e,i} &\ll \lambda_{l}^2/2c\Delta\lambda_l, \tau_I
    \label{condThomson}
\end{align}

Where $\rho_{e,i}$ is the cyclotron radius for electrons and ions, respectively, $L$ is the scattering length, $\Omega_{e,i}$ is the cyclotron frequency for electrons and ions, respectively, $\lambda_l$ is the wavelength of the laser, $\Delta \lambda_l$ is the laser linewidth, and $\tau_I$ is the integration time of the detector. Here, $\tau_I$ is 3 ns, $\lambda_l$ is $532$ nm, and $\Delta \lambda_l$ is around 0.05 nm for this experiment. The first condition is easily fulfilled by the conditions of this experiment for both the ions and electrons (i.e. the scattering length is much larger than the cyclotron radius), but the second condition is not. For the ions the left hand side gives between 0.13 and 0.89 ns and for electrons it gives between 0.25 and 1.64 $\mu$s for a full period, while the right hand side gives 9.44 ps. 

This means that the spectra detected by the CCD are the convolution between the instrumental broadening and the collective nonmagnetic TS spectra, which is given by \cite{froula}:

\begin{multline}
    S(\vec{k},\omega) = \frac{2\pi}{k} \left\lvert 1+\frac{\chi_e}{\epsilon} \right\rvert^2 f_{e}\left(\frac{\omega}{k}\right) \\
    + \frac{2\pi}{k} \sum_j \frac{Z_j^2N_j}{N_e} \left\lvert \frac{\chi_e}{\epsilon} \right\rvert^2 f_i\left(\frac{\omega}{k}\right)
    \label{thomsonForm}
\end{multline}

In particular, both spectrometers used in this work were used to measure the spectrum near the wavelength of the laser where the ion-acoustic waves (IAW) are, which corresponds to the second term in equation \ref{thomsonForm}. In order to calculate $T_e$, $T_i$, $Z$, $v_k$, and $v_{drift}$ using the IAW spectrum, a Bayesian inference model was used \cite{escalona2023}. A Markov Chain Monte Carlo (MCMC) simulation was run in order to find the posterior distribution of each parameter, and then the Bayes Theorem was used to find each parameter. An example of a fitted TS spectrum can be seen in Fig. \ref{fig:thomson1992}. Parameters that cannot be properly estimated using the IAW, such as $n_e$ and $n_i$, were obtained through the absolute calibration of the spectrometers \cite{ghazaryan2021}. Here, our vacuum chamber was filled with a background gas of $N_2$, and a laser pulse (60 mJ, 532 nm, 3ns) was used to obtain Rayleigh scattering spectra of the gas at room temperature. The Rayleigh scattering was recorded using the same optical configuration used to record the TS spectra. This process is repeated for various pressures of $N_2$ inside the chamber, which allows us to measure $n_e$ using the intensity of the TS spectrum through the following relation:

\begin{equation}
    \frac{N_T}{N_R} = \frac{k \ n_e \ \partial\sigma_T/\partial \Omega}{k \ n_{gas} \ \partial\sigma_R/\partial \Omega} = \frac{n_e \cdot 131}{n_{gas}}
    \label{rayleigh}
\end{equation}

Where $N_T$ and $N_{R}$ are the counts measured by the spectrometers for Thomson and Rayleigh Scattering respectively, $\partial\sigma_R/\partial \Omega$ and $\partial\sigma_T/\partial \Omega$ are the differential cross sections for Thomson and Rayleigh Scattering, $k$ is the probe laser wavenumber, and $n_{gas}$ is the density of the nitrogen gas used. In the case of the Andor spectrometer, $N_{R}/n_{gas}$ was found to be around $(1.2 \pm 0.3) \times 10^{-24}$ $m^3$.

A 10 mJ, 532 nm and 20 ps Nd:YAG laser (EKSPLA PL2250) was used to obtain shadowgraphy images. Before interacting with the plasma, the laser passed through a linear polarizer so that its polarization was perpendicular to the laser used in TS. After this, it passed through a plano-convex lens doublet that had a 1 mm iris placed between both lenses. From there, it went into a linear polarizer and finally into a CCD camera. Additionally, a 4-frame MCP pinhole camera was used to obtain self-emission images of the plasma, with a spatial resolution of approximately 395 $\mu$m and a 4 ns gating window per frame. These diagnostics were mainly used to measure the radius of the plasma, to monitor the instabilities present and in order to confirm that the plasma was forming properly.

\section{Results and Discussion}

\begin{figure}
    \centering
    \includegraphics[width=\linewidth]{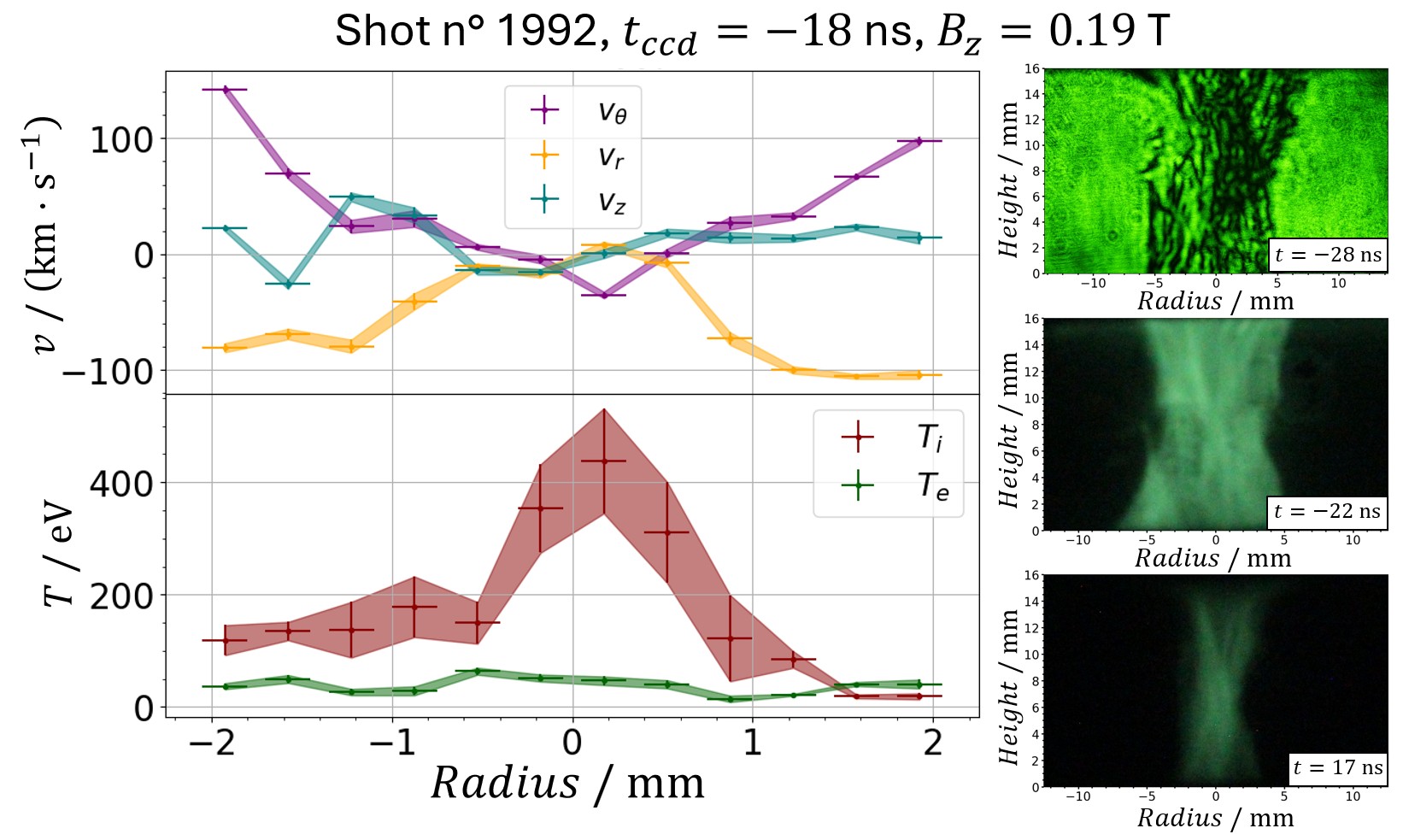}
    \caption{
    Results for shot n° 1992 (single-coil configuration, $B_{z0}=0.19$ T). The graphs on the left show the Thomson Scattering results of this shot taken at $t_{ccd}=-18$ ns: the top graph shows the spatial distribution of the radial, azimuthal and axial velocities, and the bottom graph shows the spatial distribution of the ionic and electronic temperatures. On the right, the imaging diagnostics are shown: an MCP image at $-28$ ns (top), a shadowgraphy image at $-22$ ns (middle), and a second MCP image at $17$ ns (bottom).}
    \label{1992}
\end{figure}

\begin{figure}
    \centering
    \includegraphics[width=\linewidth]{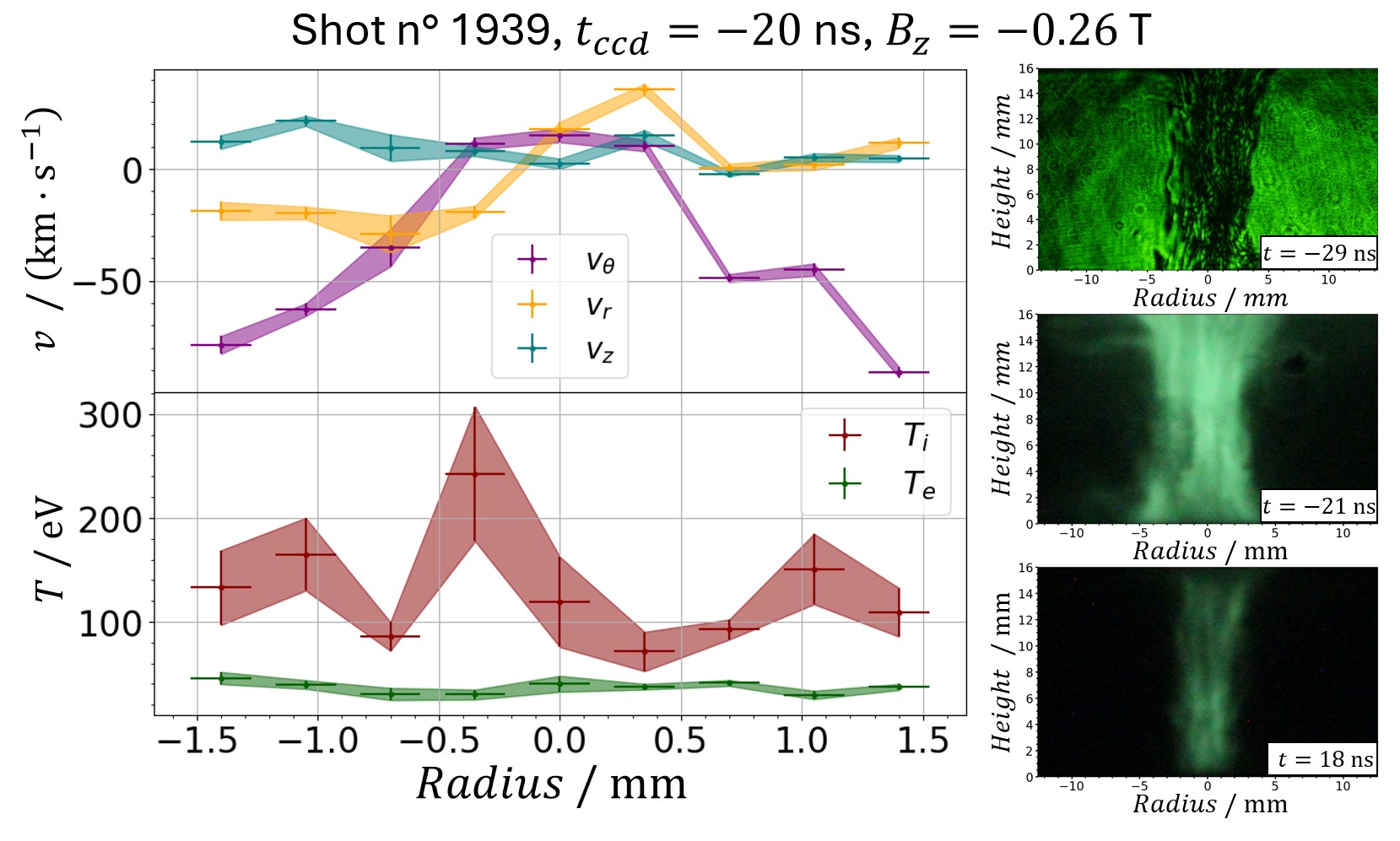}
    \caption{
    Results for shot n° 1939 (double-coil configuration, $B_{z0}=-0.26$ T). The graphs on the left show the Thomson Scattering results of this shot taken at $t_{ccd}=-20$ ns: the top graph shows the spatial distribution of the radial, azimuthal and axial velocities, and the bottom graph shows the spatial distribution of the ionic and electronic temperatures. On the right, the imaging diagnostics are shown: an MCP image at $-29$ ns (top), a shadowgraphy image at $-21$ ns (middle), and a second MCP image at $18$ ns (bottom).}
    \label{1939}
\end{figure}

An example of the effect that rotation of the plasma has on the TS spectra obtained can be seen in Fig. \ref{fig:thomson1992}, where the raw CCD spectra of shot n° 1992 are shown alongside the fitted spectra obtained by the north and south fiber bundles of the same volume of plasma. Here we see that the Doppler shift of each spectrum from 532 nm is quite different because the rotating volume of plasma is moving closer to the north fiber bundle (and thus, that spectrum is blue-shifted), and is moving farther away from the south fiber bundle (the spectrum is red-shifted). Here we can also discern visually from the raw CCD image that the relative shift between both fiber bundles gets smaller as we approach the center of the plasma. This behavior was observed consistently for most spectra obtained, from which we can conclude two things: the first one is that the effect is not limited to the periphery of the plasma, and that it has an effect throughout the diameter. And the second is that this is consistent with the idea that the effect is related to a $J_z \times B_r$ (or $J_r \times B_z$) Lorentz force rather than some other effect, since the exponential-like fall in $v_\theta$ is similar to the exponential fall that $J$ experiences because of the skin effect. In Figs. \ref{1992} and \ref{1939} we can see the spatial distribution of the azimuthal velocity, where we can see this exponential-like fall in the magnitude as we approach the center.

Similarly to what was observed in previous works \cite{cvejic2022}, in Figs. \ref{1992} and \ref{1939} we can see that the direction in which the plasma rotates changes with the polarity of the applied field. For all shots obtained in this experimental campaign, the direction of $v_\theta$ depends on the polarity of $B_{z0}$, that is, a positive $B_{z0}$ produced a clockwise rotation and a negative $B_{z0}$ produced a counter-clockwise rotation. In many cases, the direction of the azimuthal velocity changes direction near the center (see Figs. \ref{1992} and \ref{1939}, where some points close to $r=0$ mm display velocities in the opposite direction to the rest). This, rather than being something caused by a real change in behavior of the plasma, is very likely caused by the fact that at the small volume near the axis, the assumptions that were made to separate the components of the velocities are no longer valid, and therefore the measurements close to the axis are less reliable. This is due to the fact that there are many instabilities present (mainly the kink instability) that can displace the plasma, which means that the laser in many cases does not pass through the middle of the column, which can lead to the distortions in the measured velocities, especially for the probed volumes near the center of the plasma due to their proximity to the axis of rotation. And while this can be remedied mathematically if we know the distance between the center of the plasma and the center of the chamber where the laser passes through \cite{valenzuela2023}, we lack a diagnostic that can determine this distance. Therefore, given that the points close to the axis are less reliable, we can say that for all shots considered in this work, the direction of the mean azimuthal velocity is always consistent with the polarity of $B_z$, even if there are some exceptions near the axis.

\begin{figure}[t!]
    \centering
    \includegraphics[width=0.95\linewidth]{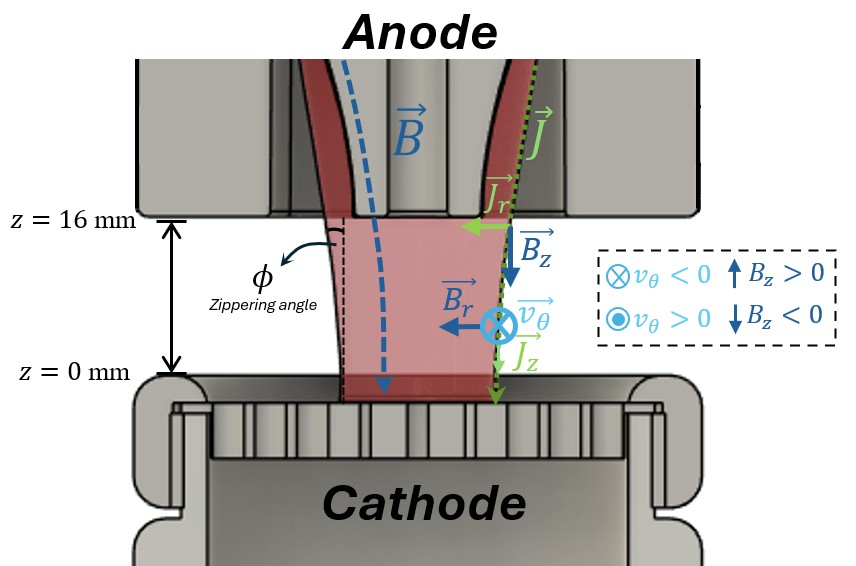}
    \caption{Diagram of the Gas-puff showing the magnetic field and the vectors being applied to the plasma when $B_{z0}$ is negative.}
    \label{fig: diagramaVector}
\end{figure}

\begin{figure*}[t!]
    \centering
    \includegraphics[width=0.95\textwidth]{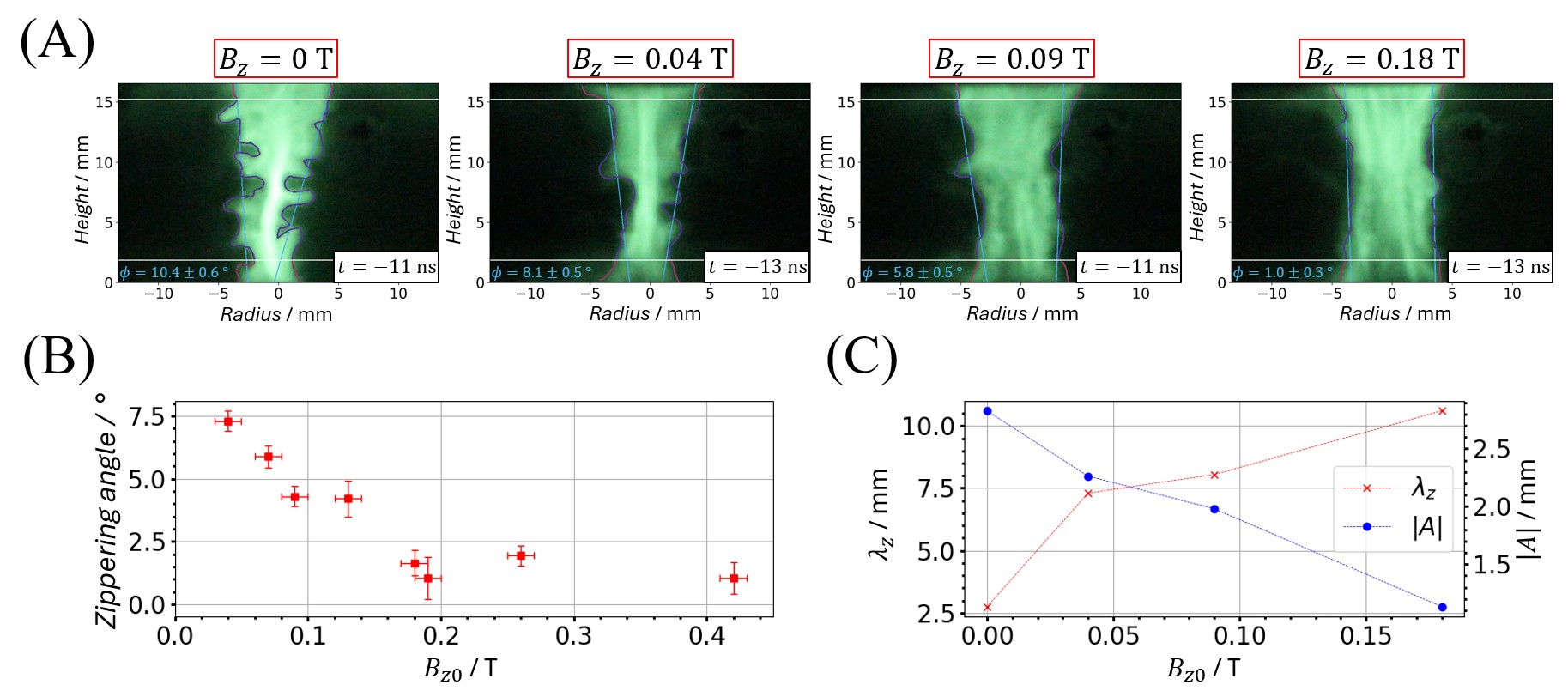}
    \caption{A) Figure of several MCP images obtained between -13 and -11 ns using different initial axial fields that shows the zippering angle of the right and left sides of each column. B) Mean zippering angle for all MCP images that were obtained in between -50 and 0 ns. C) Image showing the dominant wavelength and the amplitude of the instabilities shown in the MCP images in Fig. \ref{fig:Zipp}A.}
    \label{fig:Zipp}
\end{figure*}

Despite the fact that the direction of $v_\theta$ depends on the sign of $B_{z0}$, the plasma in this case rotates in the opposite direction that $J_r \times B_z$ points, implying that $J_z \times B_r$, which does point in the same direction in which the plasma rotates, might be the main factor in generating this effect. In Fig. \ref{fig: diagramaVector} we can see a diagram showing the imploding plasma along with the applied magnetic field and the current path, where we see that when $B_{z0}<0$, $J_r \times B_z$ points clockwise and $J_z \times B_r$ points counter-clockwise (when $B_{z0}>0$ the directions are reversed). This also seems to be the case for a previous work done on rotations in magnetized Gas-puffs \cite{cvejic2022}, where, although it was not stated explicitly, a figure similar to Fig. \ref{fig: diagramaVector} showed that the plasma rotated in the direction of $J_z \times B_r$ when nearing the stagnation phase.

In this case, for the $J_r \times B_z$ force to be large in magnitude, the zippering angle would also need to be large. This, however, did not seem to be the case for most of the MCP and shadowgraphy images taken, in which the zippering angle dropped considerably as the magnitude of the field increased (which will be discussed later in greater detail). Zippering is most often caused by an unbalanced initial density profile which causes the plasma to implode sequentially along the z-axis \cite{giuliani2015}. In our case, the Gas-puff injector's nozzle shape was designed to produce high Mach-numbers flows to suppress zippering, but residual gradients still produce a measurable inclination of the imploding plasma boundary. If we consider $\phi$ to be the zippering angle, that is, the angle formed between the vertical axis and the inclination of the visible boundary of the plasma (see Fig. \ref{fig: diagramaVector}), and assuming that $J_r$ can be calculated using $J_r\hat{r}+J_z\hat{z}=\left| J \right| (\hat{r}\sin{\phi}+\hat{z}\cos{\phi})$, then $J_r$ would decrease as the applied axial field increases. This is likely a result of the axial field stabilization effect \cite{ryutov2000}, where the counter-pressure created by the axial field as it compresses tends to even out any axial asymmetry in the implosion. On top of this, the direction in which $J_r$ pointed at the height where the TS spectra were taken ($z = 8$ mm) did not seem to have an influence on the direction of $v_\theta$, which also points to $J_r \times B_z$ not having a strong influence.

Fig. \ref{fig:Zipp}A shows several MCP images taken at similar times (between -11 and -13 ns) of several shots that used different initial axial fields. Here, we can see that as the strength of the axial field increases, the zippering angle on the plasma column gets smaller. This trend is maintained for all other cases. Fig. \ref{fig:Zipp}B shows that the average zippering angle for all shots taken between -50 and 0 ns decreases as the strength of the field increases. Additionally, in Fig. \ref{fig:Zipp}C we show the wavelength and the amplitude of the instabilities on the surface of the plasma images shown in Fig. \ref{fig:Zipp}A. Here we see that, as expected, the dominant wavelength of the MRT instabilities grows and the amplitude of the instabilities decreases as the field increases, and thus, the homogeneity of the implosion improves as the field increases.

\begin{figure}
    \centering
    \includegraphics[width=\linewidth]{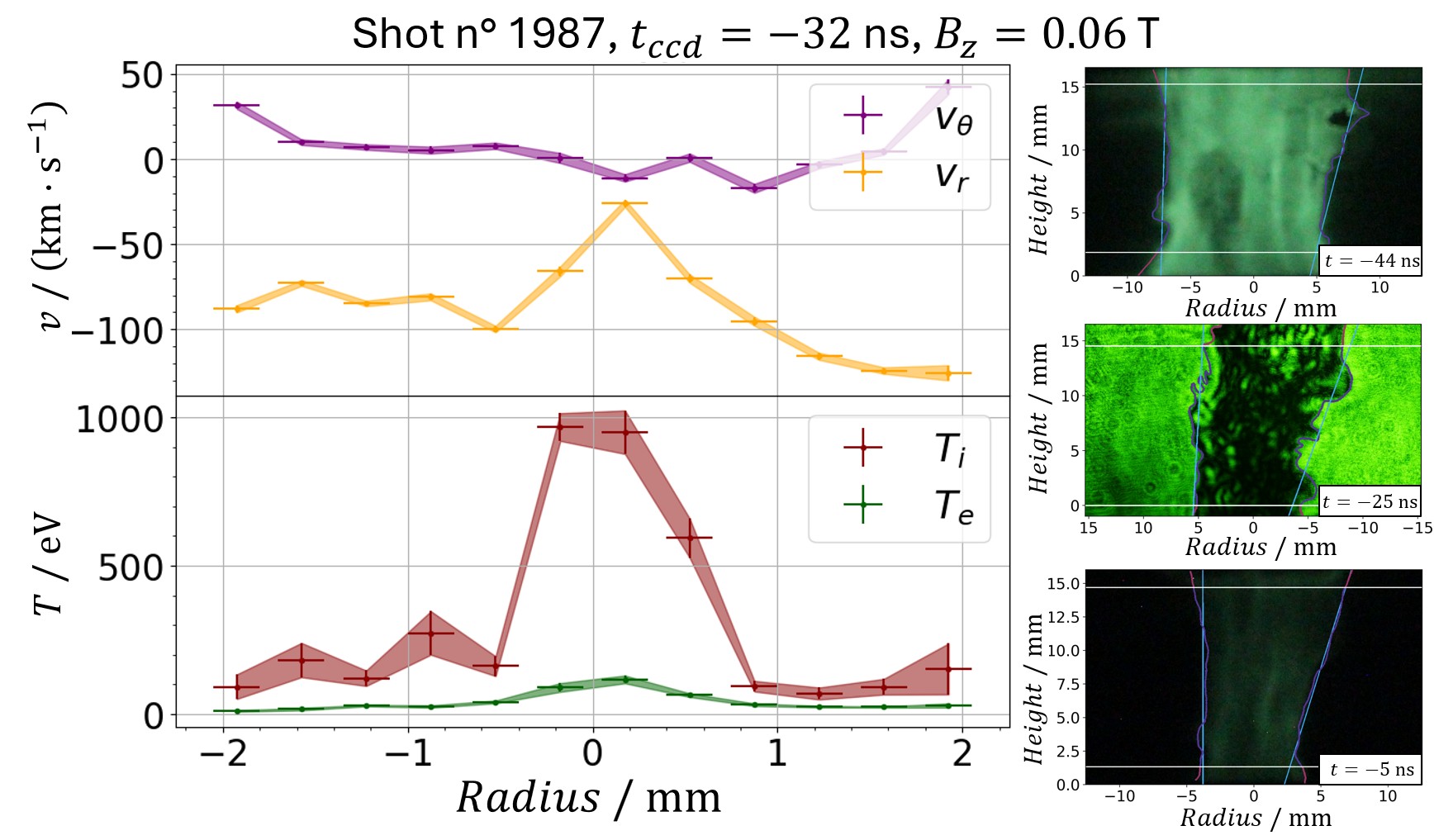}
    \caption{
    Results for shot n° 1987 (single-coil configuration, $B_{z0}=0.06$ T). The graphs on the left show the Thomson Scattering results of this shot taken at $t_{ccd}=-32$ ns: the top graph shows the spatial distribution of the radial and azimuthal velocities, and the bottom graph shows the spatial distribution of the ionic and electronic temperatures. On the right, the imaging diagnostics are shown: an MCP image at $-44$ ns (top), a shadowgraphy image at $-25$ ns (middle), and a second MCP image at $-5$ ns (bottom).}
    \label{1987}
\end{figure}

\begin{figure}
    \centering
    
    \parbox{\linewidth}{
        \centering
        \includegraphics[width=0.95\linewidth]{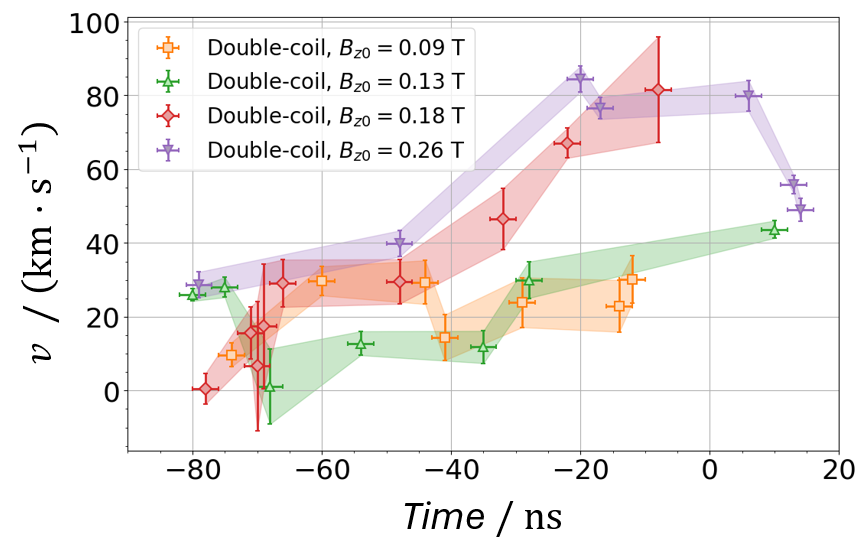}
        \par\medskip
        \textbf{(A)}
        \label{fig: vtDouble}
    }
    
    \vspace{0.3cm}
    
    \parbox{\linewidth}{
        \centering
        \includegraphics[width=0.95\linewidth]{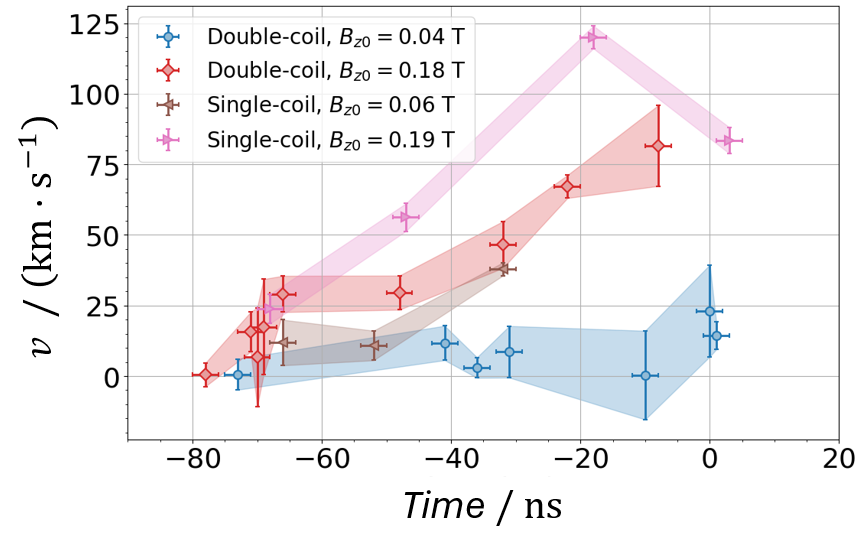}
        \par\medskip
        \textbf{(B)}
        \label{fig: vtSingle}
    }
    \caption{Figure showing the evolution of the azimuthal velocities of the outer layer of the plasma using different initial axial magnetic fields. In the case of $B_z = 0.13$ T and $0.26$ T, since some shots used a reversed polarity, we took the absolute value of the azimuthal velocity, in order to better visualize the dependence of the velocity with the magnetic field.}
    \label{fig: Vt}
\end{figure}

In shot n° 1987 (see Fig. \ref{1987}) the plasma column showed a clear inclination; it seems, however, that this did not have an influence on $v_\theta$, since the direction of the velocity did not change and the radial distribution looks relatively symmetrical, meaning that the direction of $J_r$ did not have a strong influence on the velocity. Of course, $B_r$ would also be affected by this inclination, however, since there is a relatively strong initial radial field, the direction of $B_r$ likely did not change from its initial direction at this time. All of this seems to point to $J_z \times B_r$ being the main factor that determines $v_\theta$, but in order to find conclusive evidence of the mechanism behind this phenomenon, the evolution of the radial magnetic field component needs to be measured.

\begin{figure}
    \centering
    \includegraphics[width=0.95\linewidth]{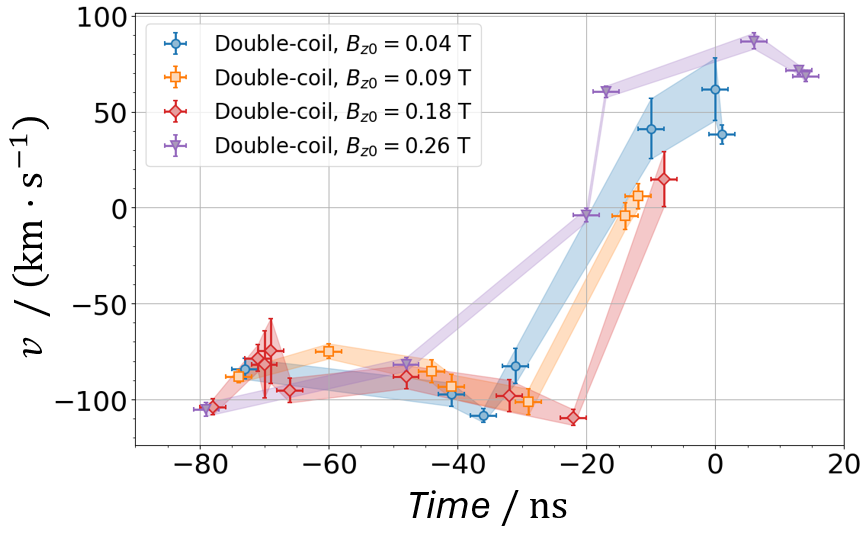}
    \caption{Figure showing the evolution of the radial velocities of the outer layer of the plasma using different initial axial magnetic fields. Some cases are not shown for the sake of clarity.}
    \label{fig: Vr}
\end{figure}

All of these results provide strong evidence that the main factor that produces the observed effects is the radial magnetic field. Regarding the double-coil case, since the initial radial field is very small, the appearance of $B_r$ could be explained by the advection of the axial field given by the non-axisymmetrical compression of the imploding plasma (as described by Fig. \ref{fig: diagramaVector}). If the radial field appears as a consequence of the compression of $B_z$, then that might explain why there is not much difference in the cases of $B_{z0}=0.18$ T and $B_{z0}=0.26$ T when looking at the time evolution of $v_{\theta}$ (see Fig. \ref{fig: Vt}A), since $B_z \cdot v_r$ would have to be dependent on $z$ for $B_r$ to appear (considering ideal MHD $\dot{\mathbf{B}}=\nabla \times (\vec{v} \times \vec{B})$). And given that zippering is fairly small and similar for both cases (see Fig. \ref{fig:Zipp}B), both $B_z$ and $v_r$ should be fairly homogeneous with respect to $z$, which means that the radial field induced in both cases could be similar. This could indicate that past a certain $B_{z0}$, given the zippering-suppressing effects of the axial field, the increase in $B_r$ through the advection of $B_z$ would begin to slow down as the applied field gets bigger. This interpretation is supported by recent measurements of the axial distribution of the radial field in a magnetized Gas-puff~\cite{cvejic2026}, which show that the value of $B_r$ can be quite large (up to 3 T), and the direction and magnitude of the field is correlated with the angle of inclination of the plasma border in the probed region. Here, near the center where the plasma is straight the magnitude of the field is smaller than in the electrodes where the plasma curves, and in the cathode the direction of $B_r$ is opposite to that in the anode, where the plasma curves in the opposite direction as well.

In Fig. \ref{fig: Vt} and \ref{fig: Vr} we can see the evolution of the radial and azimuthal velocities in the outer layer of the plasma for all the different applied fields (by 'outer layer', we mean the furthest point from the chamber axis ($r=0$ mm) at which a usable TS signal could be recorded; this corresponds to the outermost fiber that returned a fittable spectrum for each shot). The zero in the time axis represents the moment in which the stagnation happens, and it is defined by the point at which the x-ray signal measured by the diode reaches half of its peak value. We verified that the FWHM of the diode pulse does not vary appreciably with the applied axial field, so this reference is monotonic with the time of peak x-ray emission and does not introduce a $B_{z0}$-dependent bias when comparing shots.

\begin{figure}
    \centering
    \includegraphics[width=\linewidth]{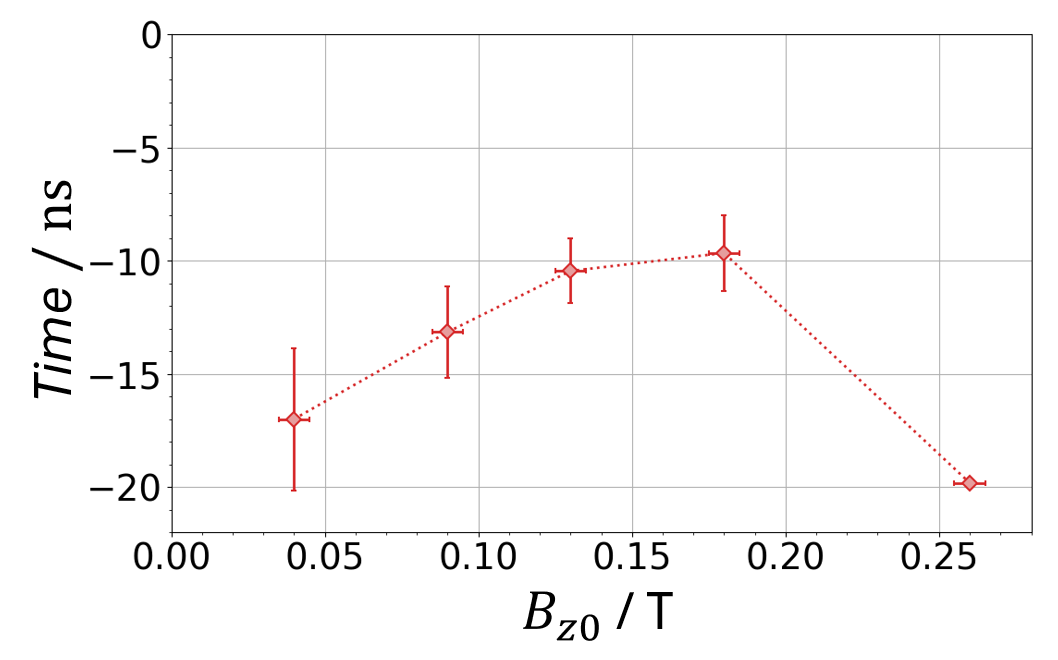}
    \caption{Time where $v_r$ reaches 0 for all cases that used the double-coil configuration}
    \label{fig: VrZ}
\end{figure}

In the case of the radial velocity, we can see that the velocity is fairly constant at 80-100 km\,s$^{-1}$ until the plasma approaches stagnation, where $v_r$ begins to change its direction. As can be seen from Fig. \ref{fig: Vr}, the change in direction happens earlier if the applied axial field is lower (not all cases are shown for better clarity, but this happens for all cases except for $B_{z0}=0.26$ T, see Fig. \ref{fig: VrZ}). This means that the time between the end of the compression and the start of x-ray emission is shortened. Considering that there does not appear to be any relation between the maximum velocity and the applied field, and that the minimum radius detected by the imaging diagnostics decreases with the applied field, we can conclude that either the implosion was delayed in the initial stages, or this delay was caused by the increased axial magnetic pressure and the centrifugal pressure increase at the border. Given that it is expected that these two forces would lengthen the implosion, as it has been measured in previous measurements \cite{rousskikh2017, mikitchuk2019}, and that the outer edges would be affected more given the distribution of $v_\theta$, it would make sense that the increase in the applied field results in this shortening of the time that the border of the plasma stops imploding and the time that the plasma starts to emit radiation.

In the case where $B_{z0}$ is $0.26$ T this trend changes, and the plasma slows down much sooner, which seems to indicate that the plasma above this point reaches a different implosion behavior regime. The reason behind this behavior is not known, but it could be related to the added axial magnetic pressure, or some other mechanism which becomes relevant when reaching a certain threshold of $B_{z0}$. Simulations of Z-pinch plasmas with sufficiently strong preembedded axial fields have shown that, under certain conditions, part of the azimuthal magnetic flux can be converted into axial flux, causing the premature stagnation of the plasma \cite{seyler2020}. The results presented here therefore could be explained by this mechanism, which leads to a much larger axial magnetic pressure and a weaker azimuthal magnetic pressure when applying stronger initial axial fields. In any case, more measurements would need to be taken near the stagnation of the plasma in order to further establish the pattern and find if this sudden change in behavior persists, or if the two points in Fig. \ref{fig: Vr} that show that the plasma slows down prematurely are outliers.

Returning to the azimuthal velocity, Fig. \ref{fig: Vt}A shows that as the applied field increases, the slope that $v_\theta$ follows also increases, and that the maximum azimuthal velocity detected is proportional to the applied field. In Fig. \ref{fig: Vt}B we see that this relation is not maintained when we compare the double-coil configuration with the single-coil configuration. In the single-coil cases of 0.06 T and 0.19 T we see that they reach higher azimuthal velocities than other cases that use the double-coil configuration, despite the fact that the axial field is similar in some cases (see for example the 0.19 T case vs the 0.18 T case, and the 0.06 T case vs the 0.04 T case in Fig. \ref{fig: Vt}A). This could be related to the initial radial magnetic field, which would be close to 0 mT in the outer layer of the plasma ($r=14$ mm, $z=8$ mm) for the double-coil configuration, and in the case of the single-coil configuration the initial radial field is 2 mT when applying a field of 0.06 T, and 8 mT when applying a field of 0.19 T at the outer layer of the injected gas.

Despite the fact that $J_r \times B_z$ points in the opposite direction that the plasma rotates into for most shots, given the semi-proportional relation between the axial field and $v_\theta$ when looking at the double-coil configuration, clearly $B_{z0}$ is important to the self-generated rotation observed. As mentioned previously, this is likely due to its importance in the evolution of $B_r$, but one alternative to this is that at the beginning of the implosion, since in some cases the plasma tends to flare out slightly as it gets closer to the cathode, $J_r \times B_z$ would point in the direction that the plasma is observed to rotate. After the implosion advances and the zippering angle inverts, this force becomes smaller and $v_\theta$ grows because of the conservation of momentum. This is unlikely however, since the specific angular momentum increases as the plasma implodes, and one would expect $J_r \times B_z$ (which now points in the opposite direction of $v_\theta$ in most cases) to be much larger after -60 ns because both $J_r$ and $B_z$ increase, and therefore the specific angular momentum should begin to fall.

The contrast between what is shown here in Fig. \ref{fig: Vt} and previous experiments where a knife-edge cathode was used is also interesting \cite{cvejic2022}, since, whereas previously $v_\theta$ remained constant throughout the implosion in the middle of the A-K gap, here we see that $v_\theta$ starts near 0 and then it starts to grow semi-linearly. This could be explained by the fact that $J_r$ is expected to be considerably smaller in the early stages of the implosion when using a round cathode, which means that $J_r \times B_z$ does not have an influence here, and that instead the effect is mainly driven by the $J_z \times B_r$ force that develops at the later stages of the implosion due to the growth of $B_r$.

We also note that MRT and kink instabilities present in the plasma can produce local variations in $J_r$ that could change the direction in which the force $J_r\times B_z$ points. However, this did not have an effect on the direction in which the plasma rotates, which suggests that $J_r\times B_z$ is comparatively small compared to $J_z\times B_r$. But, given that these instabilities are more present in the cases where $B_{z0}$ was small, it would explain why there is a greater shot-to-shot variation for $v_\theta$ when the field is small, and why $v_\theta$ does not increase much as the implosion advances (most notably for the $B_{z0}=0.09$ T case in Fig. \ref{fig: Vt}).


As for the axial velocity, the dataset was reduced relative to that of $v_r$ and $v_\theta$ by two persistent issues affecting the bottom (axial) line of sight: a lower scattered-light collection efficiency, and a faster degradation of the diagnostic window due to deposited material from the discharge. The bottom window had to be replaced approximately every ten shots to maintain calibration, while the north and south windows were replaced only at the end of an experimental run. For most of the shots where the axial velocity was able to be measured, it was fairly small and constant across the radius of the plasma (see Fig. \ref{1939} for example), or in some cases it oscillated between -20 and 20 km\,s$^{-1}$ (see Fig. \ref{1992} for example). The exception was the cases where the applied field was small ($B_{z0} < 0.1$ T), where we observed a small peak in the axial velocity near the axis of the plasma column and generally larger average axial velocities near the stagnation. The velocity points in the same direction in which the zippering progresses, meaning that the plasma is moving towards the anode from the cathode.

\begin{figure}
    \centering
    \includegraphics[width=\linewidth]{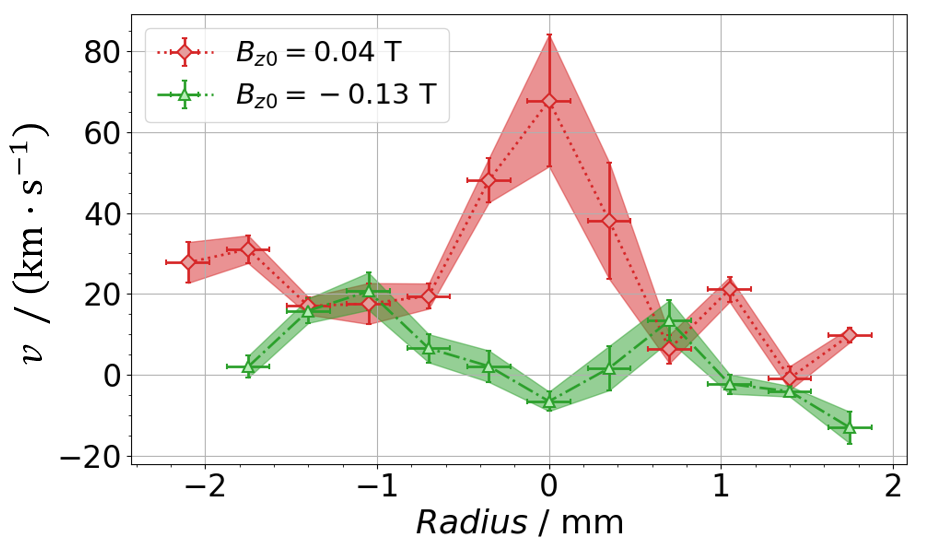}
    \caption{Radial distributions of the axial velocity for shots n° 1957 ($B_{z0}=0.04$) and n° 1927 ($B_{z0}=-0.13$) taken at $t_{ccd}=-31$ ns and $t_{ccd}=-29$ ns respectively.}
    \label{fig: VzField}
\end{figure}

In Fig. \ref{fig: VzField} we can see a comparison between the axial velocities for 2 shots taken at similar times with different initial axial fields. In the $0.04$ T case we can see that the axial velocity peaks at the center at around $68 \pm 17$ km\,s$^{-1}$, and then it falls to an average of $25 \pm 6$ km\,s$^{-1}$. In the $-0.13$ T case the velocity in the center is $-2 \pm 2$ km\,s$^{-1}$, and it does not significantly change when looking at the rest of the fibers (where the average axial velocity is $3 \pm 3$ km\,s$^{-1}$). The reason why the axial velocity behaves this way will be the subject of a future investigation, but one factor that could be related to this behavior is the zippering angle decrease caused by the increase in the applied field. In the case of Fig. \ref{fig: VzField}, there is a decrease in the zippering angle where in the case of $B_{z0}=0.04$ T the angle was $6.6$°$\pm 0.7$° and for $B_{z0}=-0.13$ T it was $3.1$°$\pm 0.4$°. This increase in the applied field could be related to the decrease in axial velocity since in simulations zippering can cause a considerable increase in the axial kinetic energy \cite{deeney1993}, and in similar plasmas with flared geometries and axial fields, the flared angle contributes to generating a pressure gradient that drives the axial velocity \cite{kumar2009}. 


\section{Conclusion}

In this paper, we have reported the first simultaneous, spatially-resolved measurement of the three velocity components in a magnetized argon gas-puff Z-pinch using Thomson Scattering along three orthogonal lines of sight. The radial, azimuthal, and axial velocities were reconstructed from the same scattering volume across the implosion, for applied axial fields between 0.04 and 0.26~T using two coil configurations. Three principal results emerge from this dataset.

\textbf{Three-component velocity measurement.} Previous Thomson-scattering studies of magnetized gas-puff Z-pinches have typically probed one or two velocity components at a time, with limited spatial coverage. The three-channel setup deployed here recovers $v_r$, $v_\theta$, and $v_z$ simultaneously from a single scattering volume, with each fiber bundle sampling a $\sim 420$ $\mu$m segment of the laser path. This capability is the basis for the two findings that follow.

\textbf{Axial velocity.} The axial velocity, which had not previously been measured experimentally in this configuration, reaches 60--70~km\,s$^{-1}$ near the axis at low applied fields ($B_{z0} < 0.1$~T) and is suppressed to within $\pm 20$~km\,s$^{-1}$ for stronger applied fields. This trend correlates with the reduction of the zippering angle, consistent with the picture that a more uniform implosion suppresses the axial pressure gradient that drives the flow~\cite{deeney1993, kumar2009}. The implication is that the axial component of the kinetic energy cannot be assumed negligible when the applied field is small, and that preembedded axial fields could improve the conversion between kinetic and thermal energy during stagnation.

\textbf{Rotation across the full plasma diameter.} The self-generated rotation is not limited to the periphery of the plasma column. The diametrical profile of $v_\theta$ measured here decreases toward the axis with an exponential-like shape consistent with the underlying current density distribution. This feature was not visible in previous Doppler-spectroscopy and edge-localized measurements~\cite{cvejic2022, Lavine2024}, where $v_\theta$ was reported only at or near the plasma boundary. Within this rotating column, the comparison between the double-coil and single-coil cases isolates the role of the seed radial field: rotation persists when the initial $B_{r0}$ at the probed plane is essentially zero, and the maximum azimuthal velocity is enhanced when a finite $B_{r0}$ is imposed. We attribute the rotation in the double-coil case to a $B_r$ that develops self-consistently during the implosion through advection and compression of the axial field, in agreement with recent direct measurements of $B_r$ in a magnetized gas-puff~\cite{cvejic2026}. Together with the qualitative observation that $v_\theta$ always follows the direction of $J_z\times B_r$ and never that of $J_r\times B_z$ across all shots in the dataset, this supports the identification of $J_z\times B_r$ as the dominant rotation driver, although direct measurement of $B_r$ and the current density distribution will be required to establish this with full certainty.

In addition to these three principal results, the radial velocity exhibits a qualitative change in the stagnation timing above $B_{z0} \approx 0.26$~T, where the deceleration begins earlier than at lower applied fields. The trend is consistent with the axial-flux-amplification mechanism reported in Hall-MHD simulations~\cite{seyler2020}, but the small number of shots in this regime in our dataset and the absence of direct $B$-field measurements preclude a definitive identification; we leave this observation as a motivation for further measurements at higher applied fields.

If the rotation is driven by a $J\times B$ Lorentz force, the bulk plasma should acquire an azimuthal flow that is, in principle, separable from the thermal motion in the ion-acoustic-wave (IAW) spectrum through the drift component of the ion velocity distribution. In our measurements, the low electron temperature ($T_e \sim 25$--$40$~eV) and the comparatively high ion temperature led to significant overlap between the IAW peaks, so the drift velocity could not be resolved with sufficient precision to provide an independent estimate of the local current density. Replicating these measurements with a heavier ion species, where the larger $Z$ would increase the IAW peak separation, would lift this limitation and is the subject of an ongoing experimental campaign. This would also allow us to measure the current density distribution, which is key to understanding the origins of the effect, and therefore it is a high priority for future experimental campaigns.

\section*{Acknowledgments}

This research was funded by the grants FONDECYT Regular n° 1220533 and 1231286, and by the grant FONDECYT Postdoctoral n° 3230401.

\section*{Author Declarations}

\subsection*{\normalfont \normalsize \textbf{Data Availability}}

The data used in this article is available from the authors upon reasonable request.

\subsection*{\normalfont \normalsize \textbf{Conflict of Interest}}

The authors have no conflicts to disclose.

\subsection*{\normalfont \normalsize \textbf{Author Contributions}}

\noindent \textbf{P. Phillips}: Conceptualization (equal); Formal analysis (lead); Investigation (equal); Methodology (equal); Data Curation (lead); Writing - Original Draft (lead); Visualization (lead). \textbf{M. Escalona}: Conceptualization (equal); Formal analysis (supporting); Investigation (equal); Methodology (equal); Writing - review \& editing (lead); Visualization (supporting); Funding acquisition (supporting). \textbf{P. Retamales}: Investigation (equal); Visualization (supporting). \textbf{M. Ribeiro}: Formal analysis (supporting); Writing - review \& editing (supporting). \textbf{F. Veloso}: Funding acquisition (supporting); Supervision (supporting); Resources (supporting); Writing - review \& editing (supporting). \textbf{J. C. Valenzuela}: Conceptualization (equal); Investigation (supporting); Writing - review \& editing (lead); Funding acquisition (lead); Project administration (lead); Resources (lead); Supervision (lead).

\bigskip


\printbibliography[title={References}]

\clearpage

\section*{Supplementary Material}

In addition to the velocity measurements included in this work, we have also included supplementary material containing shadowgraphy images of the plasma, which reveal helical and filamentary structures with unusually large pitch angles at the plasma boundary. These filaments indicate that the axial magnetic field was considerably large during the early stages of the implosion, which is directly relevant to the generation of rotations in these plasmas. These auxiliary observations, together with an order-of-magnitude estimate of the expected vs.\ observed pitch angles, are documented in the supplementary material; their detailed analysis is the subject of forthcoming work. In addition, a comparison between the TS velocity measurements and the velocity measurements obtained from the imaging diagnostics is shown in the supplementary material. The radial velocities obtained from the two methods are in good agreement.

\section*{S1. Helical and nearly-vertical filamentary structures}

The shadowgraphy diagnostic described in the main paper resolves filamentary structures at the plasma boundary. These features are not directly tied to the plasma dynamics, but they remain relevant because of what they reveal about the axial magnetic field and how the current flows in the plasma boundary. Therefore, we believe that they warrant documentation and a brief discussion. In Fig. \ref{fig:supp_Shadow130} we show and contrast these two structures (helical filaments and vertical filaments) with each other and with the unmagnetized case.

\setcounter{figure}{0}        
\renewcommand{\thefigure}{S\arabic{figure}}  

\begin{figure}
    \centering
    \includegraphics[width=\linewidth]{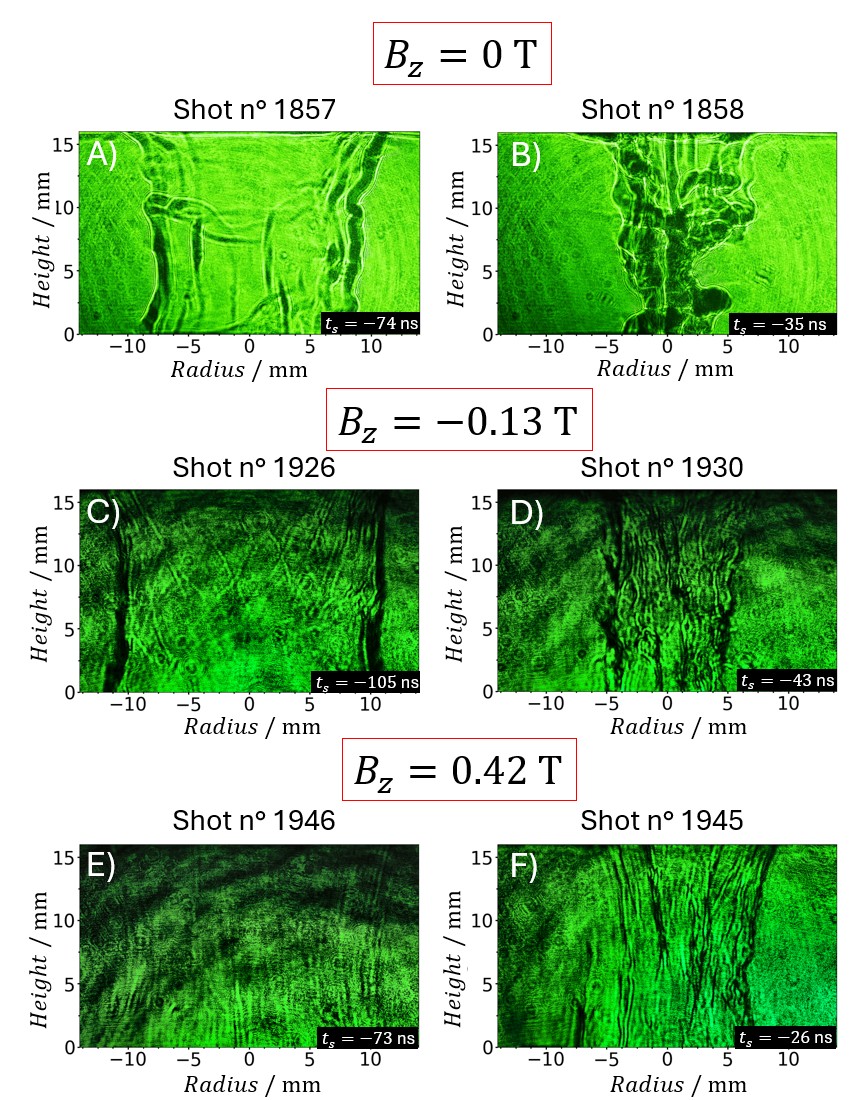}
    \caption{Shadowgraphy images of the plasma at different times using different initial axial fields. A) Shot n$^\circ$~1857 at $t=-74$~ns, $B_{z0}=0$~T. B) Shot n$^\circ$~1958 at $t=-35$~ns, $B_{z0}=0$~T. C) Shot n$^\circ$~1926 at $t=-105$~ns, $B_{z0}=-0.13$~T. D) Shot n$^\circ$~1930 at $t=-43$~ns, $B_{z0}=-0.13$~T. E) Shot n$^\circ$~1946 at $t=-73$~ns, $B_{z0}=0.42$~T. F) Shot n$^\circ$~1945 at $t=-26$~ns, $B_{z0}=0.42$~T.}
    \label{fig:supp_Shadow130}
\end{figure}

\subsection*{S1.1 Early-time helical filaments}

In these measurements, similar to other gas-puff experiments \cite{chen2024, rahman2019}, we were able to observe cross pattern shapes on the plasma, which are caused by the overlapping of the dense filaments in the back and front of the plasma, which point in different directions from the perspective of the camera and create this cross pattern. In Fig. \ref{fig:supp_Shadow130} we present several shadowgraphy images taken at early times in the pinch where we can see these helical structures. 

The pitch angle of the helical filaments remains steep ($> 60$°) throughout the compression, larger than what is reported for similar gas-puff experiments \cite{lavine2021}, which could imply that the axial magnetic field is much larger than what would be expected. As an order-of-magnitude check on this discrepancy: for shot n$^\circ$~1926 (panel~C) at the visible plasma boundary ($r\approx 14$~mm), the azimuthal field is $B_\theta \approx \mu_0 I / 2\pi r \approx 6.9$~T, while the compressed axial field is $B_z \approx -0.19$~T if we assume that the magnetic flux is conserved. The expected pitch angle, $\arctan(B_z/B_\theta) \approx 2$°, is therefore much smaller than the observed pitch angle. This would imply then that either $B_z$ is larger near the border than we would expect, or that $B_\theta$ is much smaller due to a low plasma-current coupling. Direct measurements of the axial and azimuthal magnetic fields would be needed in order to conclude anything, but these structures are certainly abnormal and could have implications for the magnitude of $B_z$ (and possibly $B_r$) and for how strong $J\times B$ is at this stage.

\subsection*{S1.2 Late-time vertical filaments}

At later times (after $-60$~ns) the helical filaments give way to nearly vertical filaments (see Fig. \ref{fig:supp_Shadow130}D), with the contrast becoming more pronounced as the applied field increases (Figs. \ref{fig:supp_Shadow130}E and \ref{fig:supp_Shadow130}F). Similar vertical structures have been reported in magnetized gas-puff experiments \cite{chen2024} and attributed to the axial field promoting gas vertical breakdown channels to the sharper parts of the mesh. The mesh geometry used in this experiment differs from those in the cited works, but we still observe these vertical structures, which could point to another cause. This effect would have to be studied in future experiments, since we do not have enough data or complementary magnetic field diagnostics to identify the cause of the effect. 

\clearpage

\section*{S2. Comparison between Thomson Scattering velocities and Imaging diagnostics}

\begin{figure}
    \centering
    \includegraphics[width=\linewidth]{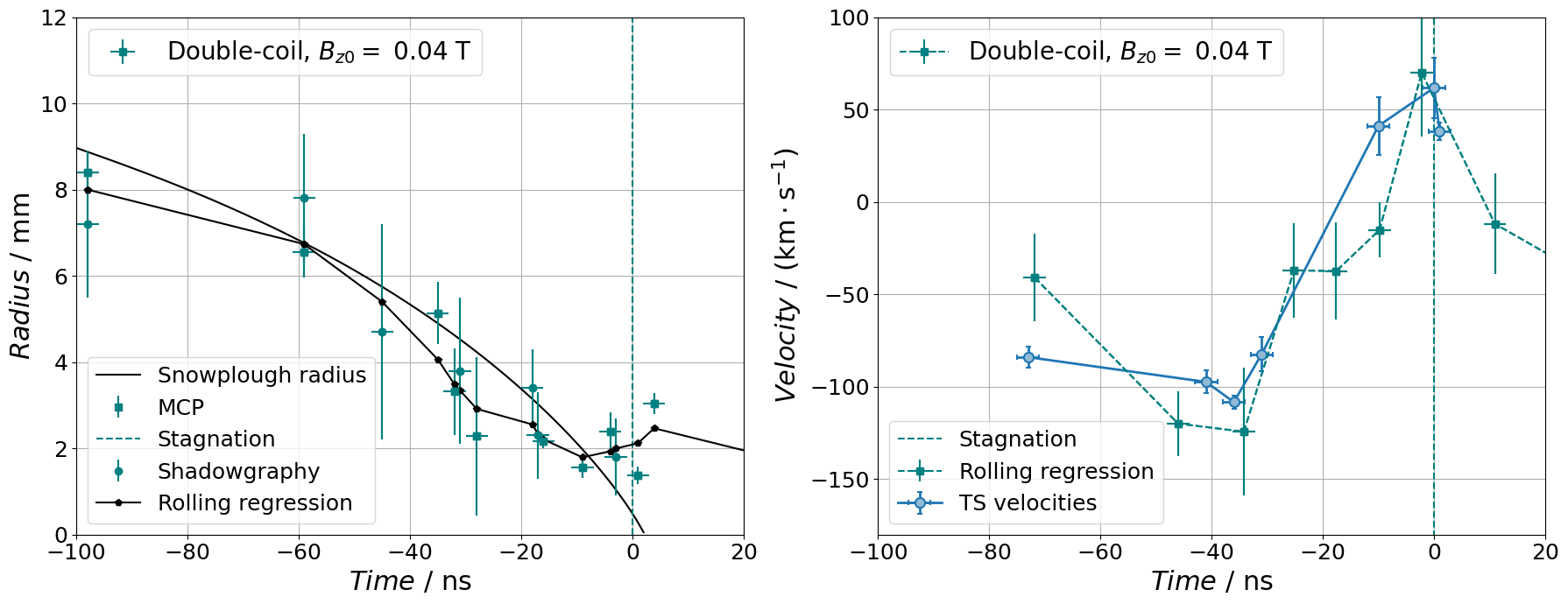}
    \caption{Image of the radius of the plasma at $z=8 mm$ obtained from the MCP images and shadowgraphy images (left), and comparison between the $v_r$ obtained using Thomson Scattering and the $v_r$ obtained from the imaging diagnostics.}
    \label{fig:vrMCP}
\end{figure}

In Fig. \ref{fig:vrMCP} we have an example of the results of the $v_r$ measured using the imaging diagnostics. The figure shows the radius measured by the MCP and shadowgraphy diagnostics. These data points were used to fit a snowplough curve \cite{giuliani2015}, and a rolling regression was also applied to obtain the mean trajectory of the plasma. This trajectory was used then to calculate the radial velocity, which showed good agreement with our Thomson Scattering measurements.

\end{document}